\documentclass[12pt,aps,prd,preprint,superscriptaddress,
   nofootinbib]{revtex4}
\newcommand{\PRE}[1]{{#1}} % Use if preprint style
\usepackage{bm} \usepackage{epsfig}
\newcommand{\mweak}{M_{\text{weak}}}

\newcommand{\mstar}{M_{*}}

\newcommand{\gev}{\text{GeV}}
\newcommand{\tev}{\text{TeV}}

\newcommand{\s}{\text{s}}
\newcommand{\yr}{\text{yr}}

\newcommand{\eqref}[1]{Eq.~(\ref{#1})}

\newcommand{\gravitino}{\tilde{G}}

\begin{document}

\preprint{UCI-TR-2004-17}  \preprint{hep-ph/0405248}

\title{
\PRE{\vspace*{1.5in}} % {\sc Essay on Gravitation} \vspace*{.4in} \\
Probing Gravitational Interactions of
Elementary Particles \PRE{\vspace*{0.3in}} }

\author{Jonathan L.~Feng\footnote{jlf@uci.edu}}
\affiliation{Department of Physics and Astronomy, University of
California, Irvine, CA 92697, USA \PRE{\vspace*{.5in}} }
\author{Arvind Rajaraman\footnote{arajaram@uci.edu}}
\affiliation{Department of Physics and Astronomy, University of
California, Irvine, CA 92697, USA \PRE{\vspace*{.5in}} }

\author{Fumihiro Takayama\footnote{fumihiro@hep.ps.uci.edu}%
\PRE{\vspace*{.2in}} } \affiliation{Department of Physics and
Astronomy, University of California, Irvine, CA 92697, USA
\PRE{\vspace*{.5in}} }

%\date{March 2004}

\begin{abstract}
\PRE{\vspace*{.3in}} The gravitational interactions of elementary
particles are suppressed by the Planck scale $\mstar \sim
10^{18}~\gev$ and are typically expected to be far too weak to be
probed by experiments.  We show that, contrary to conventional
wisdom, such interactions may be studied by particle physics
experiments in the next few years.  As an example, we consider
conventional supergravity with a stable gravitino as the lightest
supersymmetric particle.  The next-lightest supersymmetric particle
(NLSP) decays to the gravitino through gravitational interactions
after about a year. This lifetime can be measured by stopping NLSPs
at colliders and observing their decays.  Such studies will yield a
measurement of Newton's gravitational constant on unprecedentedly
small scales, shed light on dark matter, and provide a window on the
early universe. \vskip 0.5cm

{\it This essay was awarded second prize in the 2004 essay
competition of the Gravity Research Foundation.}
\end{abstract}

%\pacs{95.35.+d, 98.80.Cq, 26.35.+c, 98.80.Es}
%95.35.+d Dark matter
%98.80.Cq Particle-theory and field-theory models of the early Universe
%26.35.+c Big Bang nucleosynthesis
%98.80.Es Observational cosmology (including Hubble constant,
%          distance scale, cosmological constant, early Universe, etc)

\maketitle

As a force between elementary particles, gravity is extremely weak.
Relative to the electromagnetic, weak, and strong interactions,
gravitational interactions are suppressed by $E/\mstar$, where $E$ is
the typical energy scale of the process, and $\mstar = (8 \pi
G_N)^{-1/2} \simeq 2.4 \times 10^{18}~\gev$ is the reduced Planck
mass.  Given the energies $E \alt \tev$ accessible now and for the
foreseeable future, this is an enormous suppression.  This suppression
may be overcome in special cases, for example, in models with extra
spatial dimensions where gravity becomes strong at the TeV
scale. Barring such fortuitous scenarios, however, gravitational
effects are usually expected to be completely negligible and far
beyond the sensitivities of particle physics experiments.

In this essay, we note that this is not necessarily the case.  In
fact, in viable and well-motivated theoretical frameworks,
gravitational interactions of elementary particles may be the subject
of experimental study by the end of this decade.  Such studies may
provide new probes of gravity on the scale of elementary particles
and provide crucial insights into dark matter and early universe
cosmology.

The frameworks in which these gravitational studies may be done
include models with supersymmetry or extra dimensions in which new
particles appear at the TeV scale.  We focus first on supersymmetry, a
particularly well-motivated framework for new particle physics.
Supersymmetry predicts that each standard model particle has a
partner, its superpartner.  Supersymmetry also predicts a partner for
the graviton, the gravitino. If supersymmetry is to resolve the gauge
hierarchy problem, the standard model superpartners should have masses
around the weak scale $\mweak \sim \tev$.  A discrete symmetry,
$R$-parity, assures the stability of the lightest supersymmetric
particle (LSP) and thereby provides a dark matter candidate.  The
gauge hierarchy and dark matter problems are two of the fundamental
motivations for supersymmetry, and we assume weak-scale supersymmetry
with $R$-parity conservation below.

In supergravity models~\cite{Chamseddine:jx} where supersymmetry
breaking is mediated by the known gravitational interactions, all
superpartners, including the gravitino, have masses of the order of
$\mweak$.  The exact ordering cannot be determined theoretically.
Most studies of supergravity have assumed, either explicitly or
implicitly, that the LSP is a standard model superpartner.  Here we
explore the alternative scenario, in which the gravitino is the
LSP\footnote{The possibility of a gravitino LSP has been considered in
a number of studies, beginning with
Refs.~\cite{Pagels:ke,Weinberg:zq,Krauss:1983ik}.}.

For concreteness, let us assume that the next-lightest supersymmetric
particle (NLSP) is a charged slepton $\tilde{l}$. This slepton will be
a part of the thermal bath of the hot early universe and will freeze
out with its thermal relic density. Eventually, it will decay through
$\tilde{l} \to l \tilde{G}$. This decay is highly suppressed, as the
gravitino couples only gravitationally.  On dimensional grounds, the
lifetime of a weak-scale mass particle decaying through gravitational
interactions is $\tau \sim \mstar^2/\mweak^3 \sim \yr$.  More
precisely, the decay width is given by the
expression~\cite{Feng:2003uy}
\begin{equation}
 \Gamma(\tilde{l} \to l \tilde{G}) =
\frac{1}{48 \pi \mstar^2} \frac{m_{\tilde{l}}^5}{m_{\tilde{G}}^2}
\left[ 1 -\frac{m_{\tilde{G}}^2}{m_{\tilde{l}}^2} \right]^4 \ .
\end{equation}
In the limit $\Delta m \equiv m_{\tilde{l}} - m_{\gravitino} \ll
m_{\gravitino}$, the decay lifetime is
\begin{equation}
\tau(\tilde{l} \to l \gravitino)
 \approx 3.6 \times 10^8~\s
\left[ \frac{100~\gev}{\Delta m} \right]^4
\frac{m_{\gravitino}}{1~\tev} \ ,
\end{equation}
justifying this rough estimate.  Note that the slepton's lifetime
depends only on the NLSP and gravitino masses and Newton's constant
$G_N$, as appropriate for a gravitational decay.

It was shown recently~\cite{Feng:2003uy,Feng:2003xh,Ellis:2003dn} that
NLSP decays in the gravitino LSP scenario do not destroy the beautiful
predictions of BBN. Some combinations of NLSP and gravitino masses are
excluded, but much of the natural parameter space remains intact.  In
fact, the gravitino LSP may even have relic density
$\Omega_{\gravitino} = 0.23$ and be the dominant component of dark
matter without violating BBN constraints.  Bounds from the black-body
spectrum of the cosmic microwave background (CMB) and, in some corners
of parameter space, the diffuse photon spectrum, may be even more
severe than bounds from BBN, but these were also shown to be respected
for superpartners with weak-scale
masses~\cite{Feng:2003uy,Feng:2003xh}.  The possibility of a gravitino
LSP in supergravity theories is therefore viable.

An analogous scenario is realized if there are extra spatial
dimensions of size $\sim
\tev^{-1}$~\cite{KK,Antoniadis:1990ew,Lykken:1996fj,Appelquist:2000nn}.
Here every particle that propagates in the extra dimensions appears to
the four-dimensional observer as a tower of Kaluza-Klein (KK)
particles. For suitable models~\cite{Appelquist:2000nn}, a discrete
symmetry, KK-parity, makes the the lightest KK particle (LKP) stable.
In these scenarios, the lightest KK particles are nearly degenerate
classically.  However, quantum effects split these degeneracies. A
detailed analysis~\cite{Cheng:2002iz} shows that in many such models,
the lightest KK particle is the lightest KK graviton.  The
next-lightest KK particle (NLKP) then decays via gravitational
interactions to this graviton.  If the NLKP is a KK lepton, the decay
width is~\cite{Feng:2003uy,Feng:2003nr}
\begin{equation}
\Gamma(l^1 \to l G^1 )
 =\frac{1}{72\pi M_*^2}
\frac{m_{l^1}^7}{m_{G^1}^4} \left[1 - \frac{m_{G^1}^2}{m_{l^1}^2}
\right]^4 \left[2 + 3 \frac{m_{G^1}^2}{m_{l^1}^2} \right] \ .
\end{equation}
Again, the lifetime depends only on the parent and daughter masses
and is proportional to $G_N$.

Given the viability of the gravitino LSP scenario (and the KK
graviton LKP scenario), what can we learn?  The Large Hadron Collider
(LHC) is being built at CERN in Geneva.  In a few years, the LHC will
collide protons with protons at center-of-mass energy 14 TeV.
Weak-scale superpartners will be pair produced, and each superpartner
will rapidly decay through a chain of $R$-parity-preserving
interactions to a nearly stable NLSP.  A slepton NLSP will then
appear as a heavy charged particle passing through the collider
detector without decaying.

These sleptons are moderately relativistic and lose energy primarily
through ionization.  The softer sleptons range out within several
meters water equivalent of material and can therefore be stopped by
placing a collector just outside an LHC detector.  After a few months,
this collector may then be moved to some quiet underground environment
and monitored for slepton decays.  The slepton mass $m_{\tilde{l}}$
may be reasonably well-constrained by standard analyses of the
kinematic distributions of cascade decays. By measuring the time
distribution of decays and the energy of the outgoing leptons, one may
determine $\tau$, $m_{\tilde{G}}$, and $\mstar$, thus allowing a
calculation of $G_N$~\cite{Buchmuller:2004rq}.

The total superpartner event rate is highly dependent on the
superpartner mass spectrum.  For a light spectrum, event rates may be
as large as $10^5$ per LHC year, and a measurement of Newton's
constant at the percent level may be possible.  Such a precise
measurement of the strength of gravity between fundamental particles
will extend conventional measurements of $G_N$ to unprecedentedly
small scales, and provide a determination of $G_N$ that differs in
almost all ways from more conventional methods.

The studies described above determine $G_N$, but also have other far
reaching implications.  The gravitino's mass will be determined
simultaneously.  This implies a determination of the supersymmetry
breaking scale $F \sim m_{\gravitino} \mstar$, with implications for
models of supersymmetry breaking, the mediation of supersymmetry
breaking, and vacuum energy.  In the case of models with extra
dimensions, it may also allow us to probe the quantum effects that
produce the LKP-NLKP mass difference.

A measurement of the gravitino's mass will also have important
implications for dark matter.  If $R$-parity is conserved and the
gravitino is the LSP, gravitinos are necessarily a component of dark
matter.  Studies at future colliders will pin down supersymmetry
parameters and thereby determine the slepton thermal relic density to
high accuracy.  Along with the gravitino's mass, this will then
determine the energy density of gravitino dark matter.  If the energy
density is consistent with $\Omega_{\text{DM}} \simeq 0.23$, such
studies will have identified the main component of dark matter.  We
will also be confident that we understand the history of the universe
back to temperatures $T \sim 10~\gev$ and times $t \sim 10^{-8}~\s$,
when the slepton thermal relic density was established.

Finally, the observation of NLSP decays to gravitinos will affect our
understanding of Big Bang nucleosynthesis. While some of the parameter
space for gravitino LSPs is excluded by BBN and the CMB, much of it is
allowed, as noted above. At the boundaries of the allowed region,
deviations in BBN and CMB observations that are still consistent with
current data are predicted.  As an example, current observations of
$^7$Li are significantly lower than those predicted by standard BBN.
This anomaly can be naturally explained by $^7$Li destruction by the
late decays of NLSPs for particular decay lifetimes and NLSP relic
densities~\cite{Feng:2003uy,Feng:2003xh}. The measurement of NLSP
lifetimes and masses will therefore provide direct laboratory evidence
that will clarify our understanding of BBN and the early universe.

To conclude, we have identified well-motivated scenarios in which
particle physics experiments will be able to probe the
$\mstar$-suppressed gravitational interactions of elementary
particles. Such studies will provide new insights into gravity at
small scales, and open up a whole realm of connections between
particle physics, cosmology, and gravity.

%%%%%%%%%%%%%%%%%%%%%%%%%%%%%%%%%%%%%%%%%%%%%%%%%%%%%%

\end{document}